\documentclass[conference]{IEEEtran}
\IEEEoverridecommandlockouts

\usepackage{cite}
\usepackage{amsmath,amssymb,amsfonts}
\usepackage{algorithmic}
\usepackage{graphicx}
\usepackage{textcomp}
\usepackage{xcolor}
\usepackage{url}
\newif\ifdiffmode
\diffmodefalse 

\usepackage[normalem]{ulem}
\ifdiffmode
  \newcommand{\add}[1]{\textcolor{red}{#1}}
  \newcommand{\del}[1]{\textcolor{gray}{\sout{#1}}}
\else
  \newcommand{\add}[1]{#1}
  \newcommand{\del}[1]{}
\fi

\def\BibTeX{{\rm B\kern-.05em{\sc i\kern-.025em b}\kern-.08em
    T\kern-.1667em\lower.7ex\hbox{E}\kern-.125emX}}
\begin{document}

\title{Code-Switching Spoken Language Identification as Multi-Label Set Prediction\\

}

\author{
\IEEEauthorblockN{Shunsuke Mitsumori\textsuperscript{1},
Matthew Wiesner\textsuperscript{2},
Shigeo Morishima\textsuperscript{1},
Shinji Watanabe\textsuperscript{3}}
\IEEEauthorblockA{
\textsuperscript{1}\textit{Waseda University}, Tokyo, Japan \quad
\textsuperscript{2}\textit{Johns Hopkins University}, Baltimore, USA \quad
\\ \textsuperscript{3}\textit{Carnegie Mellon University}, Pittsburgh, USA \\
mitsumori@ruri.waseda.jp}
}

\maketitle

\begin{abstract}
Code-switched (CS) speech leaks through the monolingual language
identification (LID) filters used to curate massive speech corpora, calling
for CS-aware LID (CS-LID). We formulate utterance-level CS-LID as multi-label
language-set prediction and propose a set generator that directly outputs the
languages in an utterance, comparing it against atomic-pair and score-based
classification baselines. Oracle Top-\(k\) is the strongest baseline, but
thresholding fails because no single threshold separates CS from monolingual
speech. Our set generator predicts the correct language count on unseen pairs
without assuming the number of languages, but underperforms oracle Top-\(k\)
in exact set accuracy. 
Our analysis identifies the key obstacles to robust CS-LID: oracle cardinality, threshold instability, language bias in CS training data, and the synthetic-to-real gap.
\end{abstract}

\newcommand{\cmark}{$\surd$}
\newcommand{\dash}{--}

\section{Introduction}\label{intro}

In recent years, massive multilingual speech datasets compiled from web-crawled, in-the-wild audio~\cite{yodas,voxpopuli,emilia} have significantly advanced speech foundation models~\cite{whisper,oswmv3,oswmv4}. In the preprocessing pipelines that filter and refine language labels of such uncurated audio, utterance-level spoken language identification (LID)~\cite{voxlingua,kukk} plays a pivotal role. However, even the heavily curated YODAS~\cite{yodas} subset within OWSM v4~\cite{oswmv4} still contains code-switched (CS) speech that leaked through monolingual LID filtering~\cite{csyodas}. The problem is not the presence of CS speech itself, but the
mismatch between monolingual labels and multilingual content, which can bias
language-balanced sampling, contaminate language-specific evaluation, and
leak unintended languages into language-conditioned ASR training. This
motivates CS-aware LID (CS-LID) that identifies the set of languages present
in each utterance.

Research on code-switched speech has historically focused on a few specific language pairs (e.g., Mandarin--English or Hindi--English)~\cite{liu2025aligning,ganji2020joint,SEAME}. Recently, massively multilingual CS corpora such as CS-YODAS~\cite{csyodas} and CS-FLEURS~\cite{csflerus} have been introduced in order to explore research questions concerning how models for code-switching scale both in terms of linguistic diversity and volume of data. This scaling changes the nature of the CS-LID task: as the language inventory grows, the number of possible language pairs grows quadratically, and collecting CS training data for every pair is unrealistic.

Amidst this expansion of language coverage, existing utterance-level CS-LID approaches mainly follow one of two output strategies.
One treats each code-switched language pair as an atomic class, such as
\texttt{jpn-eng}, alongside monolingual classes~\cite{shastri2022adversarialsynthesisbaseddataaugmentation,
csyodas}. This is compatible with standard multi-class LID, but it ignores the compositional nature of code-switched speech and cannot naturally handle language pairs that were not observed during training.

Another approach keeps the output inventory as individual languages and derives a language set from LID scores. 
Patra et al.\ evaluate Top-\(k\) language predictions, using \(k=2\) for two-language CS utterances, and also fine-tune LID models with soft language-distribution targets by KL divergence
\cite{patra-etal-2026-improving}. This avoids introducing atomic pair
classes and can in principle represent unseen combinations of known languages. However, Top-\(k\) decoding assumes that the number of languages in the utterance is known in advance. Threshold-based decoding removes this explicit cardinality assumption, but its performance depends on the operating point, which can vary across languages, domains, and code-switching densities.

To overcome these limitations, we formulate utterance-level CS-LID as multi-label language-set prediction. Instead of mapping an utterance to an atomic pair label such as \texttt{jpn-eng}, our model predicts the set of individual languages present in the utterance, such as
\(\{\texttt{jpn}, \texttt{eng}\}\). This makes unseen language pairs representable as new combinations of known languages, while avoiding both the oracle language count required by Top-\(k\) inference and the operating point required by thresholding.

In this paper, we provide a controlled comparison of four representative
strategies for utterance-level CS-LID: (1) hard-target multi-class
classification with atomic pair labels, \add{(2) softmax classification over soft targets, (3) sigmoid classification over multi-hot targets, both decoded with Top-\(k\) or thresholding, and (4)} multi-label
language-set generation. We evaluate them on FLEURS, CS-FLEURS, and
real-world CS speech datasets. Atomic-pair classification performs well on seen pairs but is
structurally unable to predict unseen ones. \del{Soft-target classification}\add{Score-based classification, (2) and (3),} with Top-\(k\) is a strong oracle-cardinality baseline, while thresholding does not match Top-\(k\) accuracy. Set generation avoids \del{both }atomic pair labels\add{, oracle cardinality,} and hand-designed
thresholds, but does not yet outperform the Top-\(k\) baseline. Robust
language-set prediction for unseen, non-English, and real-world CS speech
thus remains an open challenge.
\add{Our code is publicly available at \url{https://github.com/espnet/espnet/tree/master/egs2/cs-fleurs}.}

\section{Related Work}\label{related_work}
\subsection{Frame/Token level Code-Switching Language Identification}\label{aa}
Fine-grained approaches model switching boundaries within an utterance~\cite{ganji2020joint}, exploit token- or frame-level language information for CS-ASR~\cite{liu2023reducing,dhawan2023unified,liu2025aligning}, or perform language
diarization~\cite{chua2023merlion,frost2023finetuned,liu21d_interspeech}. While useful for switch-point localization and CS-ASR, such granularity is unnecessary for corpus filtering, and local language labels are often ambiguous due to borrowings, cognates, and accent-dependent pronunciations~\cite{dogruoz2021survey,burchell2024harder}. We therefore
focus on utterance-level LID.

\subsection{Utterance level LID at Code-Switching Speech}\label{aa}
Utterance-level CS-LID has mainly followed two formulations. The first treats
each code-switched language pair as an atomic class. For example, Shastri et
al.\ classify Hindi, English, and Hindi--English as three separate
classes~\cite{shastri2022adversarialsynthesisbaseddataaugmentation}, and the
CS-YODAS baseline distinguishes \texttt{jpn} from
\texttt{jpn-eng}~\cite{csyodas}. This formulation is compatible with standard
multi-class LID, but it ties each CS label to a pair observed during
training. The second formulation keeps the output inventory as individual
languages and derives a language set from LID scores. Patra et al.\ rank
individual languages by their predicted LID probabilities and analyze the
ranks of the languages contained in CS
utterances~\cite{patra-etal-2026-improving}. They also fine-tune LID models
with soft language-distribution targets. However, deriving a set from scores either assumes the number of languages in the utterance (Top-\(k\)) or depends on an operating point that varies across languages and domains (thresholding). We address these limitations with multi-label language-set generation, as formulated in Section~\ref{sec:method}.

\section{Method}\label{sec:method}
Let $\mathcal{L}$ denote the individual-language inventory and $Y(x)\subseteq\mathcal{L}$ the reference language set for utterance $x$. We have $|Y(x)|=1$ for monolingual utterances and $|Y(x)|=2$ for two-language CS utterances. Let $\mathcal{P}_{\mathrm{train}}$ denote the language pairs observed in CS training data.

Some prior CS-LID studies treat monolingual languages and CS language pairs as mutually exclusive classes~\cite{shastri2022adversarialsynthesisbaseddataaugmentation,csyodas}. The corresponding atomic-pair class inventory is
\begin{equation}
    \mathcal{C}_{\mathrm{atomic}}
    =
    \mathcal{L}
    \cup
    \left\{
        [\ell_a,\ell_b]
        \,\middle|\,
        \{\ell_a,\ell_b\}\in\mathcal{P}_{\mathrm{train}}
    \right\},
    \label{eq:atomic-class-set}
\end{equation}
where $[\ell_a,\ell_b]$ combines two constituent languages into a single atomic label. For example, \texttt{jpn-eng} is a class distinct from both \texttt{jpn} and \texttt{eng}. The hard-target multi-class classifier predicts a single class for utterance $x$:
\begin{equation}
    \hat{c}(x)
    =
    \mathop{\mathrm{arg\,max}}_{c\in\mathcal{C}_{\mathrm{atomic}}}
    p_{\mathrm{atomic}}(c\mid x),
    \label{eq:atomic-prediction}
\end{equation}
where $p_{\mathrm{atomic}}(c\mid x)$ is the probability assigned to class $c$. However, if a test-time pair satisfies $\{\ell_a,\ell_b\}\notin\mathcal{P}_{\mathrm{train}}$, the corresponding class is absent, so the classifier cannot predict the pair even if both languages have been learned individually.

One way to avoid this limitation is to score individual languages and apply Top-\(k\) decoding. Given the reference language count $k=|Y(x)|$, this returns the $k$ languages with the highest logits $z_\ell(x)$. For two-language CS utterances, it becomes
\begin{equation}
    \hat{Y}_{\mathrm{top2}}(x)
    =
    \operatorname{Top2}_{\ell\in\mathcal{L}}z_\ell(x).
    \label{eq:top2-prediction}
\end{equation}
Patra et al.~\cite{patra-etal-2026-improving} analyze the ranks of the languages in CS utterances using predicted LID probabilities. Top-2 permits unseen combinations of known languages without pair-specific classes. In the oracle-cardinality setting, however, $k=|Y(x)|$ is supplied rather than predicted.

Patra et al.~\cite{patra-etal-2026-improving} also adapt LID models to CS speech using a KL divergence objective. Our soft-target classifier follows this approach. It uses one-hot targets for monolingual utterances and, for CS utterances, a target distribution that assigns equal mass to the constituent languages:
\begin{equation}
    q(\ell\mid x)=
    \begin{cases}
        1/|Y(x)|, & \ell\in Y(x),\\
        0, & \ell\notin Y(x).
    \end{cases}
    \label{eq:soft-target}
\end{equation}
The model minimizes
\begin{equation}
    \mathcal{J}_{\mathrm{KL}}
    =
    D_{\mathrm{KL}}
    \bigl(q(\cdot\mid x)\|p_\theta(\cdot\mid x)\bigr),
    \label{eq:kl-loss}
\end{equation}
where $p_\theta(\cdot\mid x)$ is the predicted softmax distribution over individual languages.

We also include a sigmoid multi-label classifier as a baseline. Instead of a softmax distribution, it predicts a sigmoid score $r_\ell(x)=\sigma(z_\ell(x))$ for each language. It uses binary multi-hot targets, with $y_\ell(x)=1$ for $\ell\in Y(x)$ and $0$ otherwise, and is trained with binary cross-entropy (BCE).

Both score-based classifiers support oracle Top-\(k\) decoding or thresholding on their softmax probabilities or sigmoid scores. Thresholding removes the need for oracle cardinality, but requires choosing a score threshold.

To avoid both oracle cardinality and score thresholds, we use a multi-label set generator that predicts both how many and which languages are present. It directly generates a language-token sequence $\hat{s}(x)$, which is converted into an unordered language set:
\begin{equation}
    \hat{Y}(x)
    =
    \operatorname{set}\bigl(\hat{s}(x)\bigr)\cap\mathcal{L}.
    \label{eq:predicted-language-set}
\end{equation}
For example, generating \texttt{<jpn>} yields $\hat{Y}(x)=\{\mathrm{jpn}\}$, and \texttt{<jpn> <eng>} yields $\{\mathrm{jpn},\mathrm{eng}\}$. The predicted language count is $\hat{k}(x)=|\hat{Y}(x)|$. An utterance is regarded as monolingual when $\hat{k}(x)=1$ and as CS when $\hat{k}(x)\geq2$. This removes the need for a supplied language count and a score threshold, and makes unseen pairs representable as combinations of known languages.

\section{Experiments}\label{3}
\subsection{Datasets}\label{3a}
In this paper, we conduct our experiments using five speech datasets. All language labels are represented using ISO 639-3 language codes, such as \texttt{eng}, \texttt{jpn}, and \texttt{cmn}. Specifically, FLEURS~\cite{flerus} and CS-FLEURS~\cite{csflerus} are used for both training and evaluation. CS-YODAS~\cite{csyodas} is utilized exclusively for training, while the Bangor Miami Corpus~\cite{miami} and the Multilingual Soap Opera Corpus~\cite{soap} are reserved solely for evaluation.

\begin{table}[t]
\centering
\caption{Matrix and embedded language inventories of CS-FLEURS subsets used in this work. Language codes follow ISO 639-3.}
\vspace{-2mm}
\label{tab:csfleurs_lang_inventory}
\scriptsize
\setlength{\tabcolsep}{2.0pt}
\renewcommand{\arraystretch}{1.15}
\begin{tabular}{|p{0.15\columnwidth}|p{0.39\columnwidth}|p{0.39\columnwidth}|}
\hline
\textbf{Subset} & \textbf{Matrix languages} & \textbf{Embedded languages} \\
\hline
Read
& {\ttfamily ara deu fra hin por rus\newline
spa ces cmn ita jpn kor\newline
slk tel}
& {\ttfamily eng} \\
\hline

XTTS1
& {\ttfamily ara deu fra hin por rus\newline
spa ces cmn ita jpn kor\newline
hun nld pol tur}
& {\ttfamily eng} \\
\hline

XTTS2
& {\ttfamily ara cmn hin spa}
& {\ttfamily ara fra deu ita por pol\newline
tur rus nld ces spa cmn\newline
jpn hun kor hin} \\
\hline

MMS
& {\ttfamily ara deu fra hin por rus\newline
spa nld pol tur ben bul\newline
cat ceb cym ell fin guj\newline
heb hun ind isl jav kan\newline
kaz kir lav lug mal mar\newline
mya pan ron swe swh tam\newline
tel tgk tgl tha ukr urd\newline
uzb yor zlm}
& {\ttfamily eng} \\
\hline
\end{tabular}
\vspace{-2mm}
\end{table}

\textbf{FLEURS}~\cite{flerus} is a massively multilingual speech dataset built from FLoRes text and covers 102 languages. Each utterance is associated with a single language label. We use FLEURS as monolingual LID data for both training and evaluation. We use the official train and test splits of FLEURS.

\textbf{CS-FLEURS}~\cite{csflerus} is a massively multilingual code-switched speech dataset constructed
from FLEURS-derived text. It contains both human-read and synthetic code-switched speech using XTTS or MMS-TTS~\cite{xtts,mms-tts}. The training subset is synthesized entirely using XTTS, while the test data is categorized into four distinct subsets: Read, XTTS1, XTTS2, and MMS. Table~\ref{tab:csfleurs_lang_inventory} summarizes the language pairs included in each subset. In this table, the matrix language denotes the dominant language of the utterance, while the embedded language is the one inserted into it. The READ and MMS test sets contain both language pairs seen during training and unseen pairs. In contrast, the XTTS1 subset consists exclusively of seen language pairs, whereas the XTTS2 subset is composed entirely of unseen pairs.
In our experiments, we conduct a fine-grained evaluation by explicitly partitioning the results into seen and unseen conditions across these subsets.

\begin{table*}[!t]
\centering
\caption{Summary of model configurations. Check marks indicate the datasets used for training.}
\label{tab:model_configs}
\scriptsize
\setlength{\tabcolsep}{3.0pt}
\renewcommand{\arraystretch}{1.12}
\begin{tabular}{l l c c c p{0.30\textwidth}}
\hline
\textbf{ID} & \textbf{Model family} & \textbf{FLEURS} & \textbf{CS-FLEURS} & \textbf{CS-YODAS} & \textbf{Target form} \\
\hline
H1 & Hard-target multi-class classifier
   & \cmark & \dash & \dash
   & Individual language class \\

H2 & Hard-target multi-class classifier
   & \cmark & \cmark & \dash
   & Individual language class / atomic pair class \\

H3 & Hard-target multi-class classifier
   & \cmark & \cmark & \cmark
   & Individual language class / atomic pair class \\

S1 & Soft-target classifier
   & \cmark & \cmark & \dash
   & Soft distribution over individual languages \\

S2 & Soft-target classifier
   & \cmark & \cmark & \cmark
   & Soft distribution over individual languages \\

M1 & Sigmoid multi-label classifier
   & \cmark & \cmark & \dash
   & Binary multi-hot vector over individual languages \\

M2 & Sigmoid multi-label classifier
   & \cmark & \cmark & \cmark
   & Binary multi-hot vector over individual languages \\

G1 & Multi-label set generator (ours)
   & \cmark & \cmark & \dash
   & Unordered language-token set \\

G2 & Multi-label set generator (ours)
   & \cmark & \cmark & \cmark
   & Unordered language-token set \\
\hline
\end{tabular}
\vspace{-2mm}
\end{table*}

\textbf{CS-YODAS}~\cite{csyodas} consists of naturally occurring CS speech mined from YouTube. Following the paper, we select language pairs combining English with Arabic, Mandarin, French, Hindi, Japanese, and Russian. \del{This dataset is incorporated to investigate whether in-the-wild supervision enhances robustness beyond controlled settings for trilingual CS utterances.}\add{We use it to test whether in-the-wild supervision improves robustness beyond controlled settings.}

\textbf{The Bangor Miami Corpus}~\cite{miami} is a conversational Spanish--English bilingual corpus. We leverage this dataset to analyze how model performance shifts as the proportion of English within a CS utterance varies.

\textbf{Multilingual Soap Opera Corpus}~\cite{soap} features South African broadcast speech with naturally occurring CS among English, isiXhosa, and isiZulu. This corpus serves to test whether CS-LID models can scale to three-language CS scenarios.

\begin{table*}[t]
\centering
\caption{Exact set accuracy (\%) on FLEURS and CS-FLEURS. FL, CS-FL, and CS-YO denote the training datasets FLEURS, CS-FLEURS, and CS-YODAS. $\dagger$ marks oracle Top-\(k\) ($k=1$ for FLEURS and $k=2$ for CS-FLEURS). \#Params excludes the shared 965M-parameter MMS-1B frontend. ``--'' means not applicable. Bold marks column maxima, including oracle results.}
\vspace{-2mm}
\label{tab:overall_results}
\scriptsize
\setlength{\tabcolsep}{3.8pt}
\renewcommand{\arraystretch}{1.12}
\begin{tabular}{@{}llccccrrrr@{}}
\hline
\textbf{ID} & \textbf{Model family} & \textbf{\#Params}
& \textbf{FL} & \textbf{CS-FL} & \textbf{CS-YO}
& \multicolumn{1}{c}{\textbf{FLEURS}}
& \multicolumn{3}{c}{\textbf{CS-FLEURS}} \\
\cline{7-7} \cline{8-10}
& & & & &
& \textbf{All}
& \textbf{All} & \textbf{Seen} & \textbf{Unseen} \\
\hline
H1-Top1$^\dagger$
& Hard-target multi-class classifier
& 9.33M
& \cmark & \dash & \dash
& 96.7
& \dash & \dash & \dash \\

H1-Top2$^\dagger$
& Hard-target multi-class classifier
& 9.33M
& \cmark & \dash & \dash
& \dash
& 8.3 & 10.7 & 6.7 \\

S1-Top1$^\dagger$
& Soft-target classifier
& 9.32M
& \cmark & \cmark & \dash
& 91.4
& \dash & \dash & \dash \\

S1-Top2$^\dagger$
& Soft-target classifier
& 9.32M
& \cmark & \cmark & \dash
& \dash
& 52.0 & 95.2 & 22.2 \\

S2-Top1$^\dagger$
& Soft-target classifier
& 9.32M
& \cmark & \cmark & \cmark
& 88.2
& \dash & \dash & \dash \\

S2-Top2$^\dagger$
& Soft-target classifier
& 9.32M
& \cmark & \cmark & \cmark
& \dash
& 33.0 & 59.7 & 14.5 \\

M1-Top1$^\dagger$
& Sigmoid multi-label classifier
& 9.32M
& \cmark & \cmark & \dash
& 96.5
& \dash & \dash & \dash \\

M1-Top2$^\dagger$
& Sigmoid multi-label classifier
& 9.32M
& \cmark & \cmark & \dash
& \dash
& \textbf{59.7} & 96.6 & \textbf{34.2} \\

M2-Top1$^\dagger$
& Sigmoid multi-label classifier
& 9.32M
& \cmark & \cmark & \cmark
& 93.6
& \dash & \dash & \dash \\

M2-Top2$^\dagger$
& Sigmoid multi-label classifier
& 9.32M
& \cmark & \cmark & \cmark
& \dash
& 58.7 & \textbf{96.8} & 32.4 \\
\hline
H2
& Hard-target multi-class classifier
& 9.33M
& \cmark & \cmark & \dash
& 97.3
& 30.0 & 73.5 & 0.0 \\

H3
& Hard-target multi-class classifier
& 9.33M
& \cmark & \cmark & \cmark
& \textbf{97.4}
& 30.1 & 73.7 & 0.0 \\

G1
& Multi-label set generator (ours)
& 23.55M
& \cmark & \cmark & \dash
& 95.6
& 36.6 & 73.5 & 11.1 \\

G2
& Multi-label set generator (ours)
& 23.55M
& \cmark & \cmark & \cmark
& 93.1
& 30.1 & 73.5 & 0.0 \\
\hline
\end{tabular}
\vspace{-2mm}
\end{table*}

\subsection{Models}\label{3b}

All systems use the same 965M-parameter MMS-1B frontend, which is also updated
during training: its multilayer representations are combined via a learnable
weighted sum and normalized at the utterance level. Excluding MMS-1B, the trainable
parameters are 9.33M, 9.32M, 9.32M, and 23.55M for the hard-target, soft-target,
sigmoid, and set-generator models, so the largest gap accounts for less than 2\% of
the parameters updated during training.

The three classifiers are based on the ESPnet LID recipe and share the same
architecture and optimization, differing only in the output head and the objective
defined in Section~\ref{sec:method}. The frontend features are passed through an
ECAPA-TDNN encoder, channel-attentive statistics pooling, and a RawNet3-style
projector to obtain an utterance-level embedding, on which a subcenter cosine head
with $K=3$ operates, with an additive angular margin of $m=0.5$ and scale $s=30$
applied to the positive languages. Each model is trained for 3 epochs with an
effective batch size of 32; the learning rate is linearly warmed up from
$3\times10^{-6}$ to $5\times10^{-6}$ over the first 10\% of the total steps, held
constant for the next 20\%, and linearly decayed to $5\times10^{-7}$.

\paragraph{Hard-target multi-class classifier}
Trained without CS data, the model has no CS class, so we take the top-2 individual
languages for CS utterances; trained with CS data, the predicted atomic pair is
converted back to an unordered language set.

\paragraph{Soft-target classifier}
CS utterances are trained with soft targets assigning 0.5 probability mass to each
constituent language.

\paragraph{Sigmoid-based multi-label classifier}
The BCE objective uses a positive-class weight of 50. 
At inference, both score-based classifiers take the top-$k$ languages for oracle evaluation, where $k$ is the reference language count; we additionally sweep the score threshold.

\paragraph{Multi-label set generator}
The set generator is based on the ESPnet ASR recipe. The frontend features are
passed to a 24-layer Transformer encoder, and a 4-layer Transformer decoder with 4
attention heads and 256 hidden dimensions generates the language tokens. It is
trained with decoder cross-entropy only, without CTC loss; since the task is
unordered, we compute the loss for both token orders of each CS utterance and take
the smaller one.

Unlike the three classifiers, the set generator uses two-stage training with the same data and objective in both stages. The first stage stabilizes early Transformer training: MMS-1B is frozen and only the
randomly initialized encoder-decoder is updated, about 1/40 of the parameters
trained in the second stage. We then initialize from the best validation-loss
checkpoint and unfreeze MMS-1B. Each stage runs for 3 epochs, with a peak learning
rate of $10^{-3}$ and $5\times10^{-6}$ respectively; in both, the learning rate is
linearly warmed up from near zero over the first 10\% of the total steps and then
decayed by an inverse square-root schedule.

\subsection{Experimental Configurations}
\label{sec:experimental_configurations}
We train nine configurations, summarized in Table~\ref{tab:model_configs}. They differ in model family and in whether CS-FLEURS and CS-YODAS are included in training.

We report exact set accuracy: a prediction is correct only if $\hat{Y}(x)=Y(x)$, regardless of language order. For CS-FLEURS, seen and unseen pairs are distinguished by whether the pair appears in CS training data. Unseen pairs consist of languages individually observed in training, but not in that combination.


\begin{table*}[t]
\centering
\caption{Subset-wise exact set accuracy (\%) on CS-FLEURS. Seen and Unseen denote whether each language pair appears in CS training data. $\dagger$ marks oracle Top-2. Other notation follows Table~\ref{tab:overall_results}.}
\vspace{-2mm}
\label{tab:csfleurs_subset}
\scriptsize
\setlength{\tabcolsep}{2.2pt}
\renewcommand{\arraystretch}{1.12}
\begin{tabular}{@{}llccccrrrrrr@{}}
\hline
\textbf{ID} & \textbf{Model family} & \textbf{\#Params}
& \textbf{FL} & \textbf{CS-FL} & \textbf{CS-YO}
& \multicolumn{2}{c}{\textbf{Read}}
& \textbf{XTTS1}
& \textbf{XTTS2}
& \multicolumn{2}{c}{\textbf{MMS}} \\
\cline{7-8} \cline{9-9} \cline{10-10} \cline{11-12}
& & & & &
& \textbf{Seen} & \textbf{Unseen}
& \textbf{Seen}
& \textbf{Unseen}
& \textbf{Seen} & \textbf{Unseen} \\
\hline
H1-Top2$^\dagger$
& Hard-target multi-class classifier
& 9.33M
& \cmark & \dash & \dash
& 14.6 & 1.2
& 1.7
& \textbf{0.5}
& 33.7 & 16.3 \\

S1-Top2$^\dagger$
& Soft-target classifier
& 9.32M
& \cmark & \cmark & \dash
& 82.7 & 1.6
& \textbf{99.7}
& 0.0
& 99.8 & 56.7 \\

S2-Top2$^\dagger$
& Soft-target classifier
& 9.32M
& \cmark & \cmark & \cmark
& 5.7 & 1.0
& 77.2
& 0.0
& 84.5 & 37.0 \\

M1-Top2$^\dagger$
& Sigmoid multi-label classifier
& 9.32M
& \cmark & \cmark & \dash
& 88.1 & 1.2
& 99.6
& 0.0
& \textbf{99.9} & \textbf{87.3} \\

M2-Top2$^\dagger$
& Sigmoid multi-label classifier
& 9.32M
& \cmark & \cmark & \cmark
& \textbf{88.6} & \textbf{19.0}
& 99.6
& 0.0
& \textbf{99.9} & 82.0 \\
\hline
H2
& Hard-target multi-class classifier
& 9.33M
& \cmark & \cmark & \dash
& 0.2 & 0.0
& 99.6
& 0.0
& \textbf{99.9} & 0.0 \\

H3
& Hard-target multi-class classifier
& 9.33M
& \cmark & \cmark & \cmark
& 0.8 & 0.0
& \textbf{99.7}
& 0.0
& \textbf{99.9} & 0.0 \\

G1
& Multi-label set generator (ours)
& 23.55M
& \cmark & \cmark & \dash
& 0.0 & 0.0
& \textbf{99.7}
& 0.0
& 99.8 & 28.5 \\

G2
& Multi-label set generator (ours)
& 23.55M
& \cmark & \cmark & \cmark
& 0.2 & 0.0
& \textbf{99.7}
& 0.0
& 99.6 & 0.0 \\
\hline
\end{tabular}
\vspace{-2mm}
\end{table*}

\begin{table*}[t]
\centering
\caption{Two-language prediction rate (\%) on CS-FLEURS. For hard-target
classifiers, atomic-pair predictions count as two-language predictions; for
set generators, the rate is computed from unique generated language tokens.
Notation follows Table~\ref{tab:overall_results}.}
\vspace{-2mm}
\label{tab:pred_cardinality}
\scriptsize
\setlength{\tabcolsep}{2.0pt}
\renewcommand{\arraystretch}{1.12}
\begin{tabular}{@{}llccccrrrrrr@{}}
\hline
\textbf{ID} & \textbf{Model family} & \textbf{\#Params}
& \textbf{FL} & \textbf{CS-FL} & \textbf{CS-YO}
& \shortstack{\textbf{Read}\\\textbf{Seen}}
& \shortstack{\textbf{Read}\\\textbf{Unseen}}
& \shortstack{\textbf{XTTS1}\\\textbf{Seen}}
& \shortstack{\textbf{XTTS2}\\\textbf{Unseen}}
& \shortstack{\textbf{MMS}\\\textbf{Seen}}
& \shortstack{\textbf{MMS}\\\textbf{Unseen}} \\
\hline
H2
& Hard-target multi-class classifier
& 9.33M
& \cmark & \cmark & \dash
& 0.2
& 0.0
& 99.9
& 99.9
& \textbf{99.9}
& 38.5 \\

H3
& Hard-target multi-class classifier
& 9.33M
& \cmark & \cmark & \cmark
& \textbf{0.8}
& 0.0
& 99.8
& 99.7
& \textbf{99.9}
& 14.6 \\

G1
& Multi-label set generator (ours)
& 23.55M
& \cmark & \cmark & \dash
& 0.0
& 0.0
& \textbf{100.0}
& \textbf{100.0}
& \textbf{99.9}
& \textbf{96.8} \\

G2
& Multi-label set generator (ours)
& 23.55M
& \cmark & \cmark & \cmark
& 0.2
& 0.0
& \textbf{100.0}
& 99.9
& 99.7
& 43.1 \\
\hline
\end{tabular}
\end{table*}

\section{Results}
\label{sec:results}

This section presents the evaluation results of the \del{four models} \add{nine configurations} from Section~\ref{sec:experimental_configurations} on FLEURS, CS-FLEURS, the Bangor Miami Corpus, and the Multilingual Soap Opera Corpus. Notably, our Top-$k$ decoding evaluation relies on prior knowledge of oracle cardinality ($k=1$ for monolingual FLEURS and $k=2$ for CS-FLEURS). Because the multi-label set generator solves a strictly harder problem by predicting a variable-length language sequence without such prior knowledge, direct performance comparisons do not necessarily reflect architectural superiority.

\paragraph{Overall results}
Table~\ref{tab:overall_results} summarizes the results on FLEURS and CS-FLEURS. For CS-FLEURS, we separately report results on seen and unseen language pairs.

On FLEURS, the hard-target classifiers maintain high monolingual LID accuracy. H1-Top1, H2, and H3 achieve 96.7\%, 97.3\%, and 97.4\%, respectively. The multi-label set generator G1 also remains competitive, achieving 95.6\%. In contrast, the soft-target classifier S1-Top1 achieves a lower monolingual accuracy of 91.4\%. This suggests a conflict when training a single softmax model with both one-hot and soft targets: monolingual examples encourage probabilities to concentrate on one language, whereas CS examples distribute them across two. \add{The sigmoid classifier M1-Top1 avoids this conflict and reaches 96.5\%, since its per-language scores are not normalized to sum to one.}

On CS-FLEURS, the hard-target multi-class classifier H2 achieves high accuracy on seen pairs but obtains 0.0\% on unseen pairs. This confirms the structural limitation of the atomic-pair formulation: an unseen language pair has no corresponding output class. The FLEURS-only baseline H1-Top2 can theoretically predict unseen combinations, but its accuracy remains low.
\del{The soft-target classifier with Top-2 decoding, S1-Top2, is the strongest Top-k baseline for CS utterances. This shows that KL-based soft-target training over individual language classes is a strong alternative to atomic-pair classification when oracle cardinality is available.}
\add{The score-based classifiers with Top-2 decoding are the strongest baselines for CS utterances, led by the sigmoid classifiers M1-Top2 and M2-Top2 at 59.7\% and 58.7\% overall. This shows that classifying over individual languages, with targets that mark all constituent languages of an utterance, is a strong alternative to atomic-pair classification when oracle cardinality is available.}

With the same training data, G1 improves unseen accuracy to 11.1\%, compared to 0.0\% for H2. This indicates that language-token set generation alleviates the structural limitation of atomic-pair classification. However, the current implementation remains below \del{S1-Top2}\add{both score-based baselines with oracle cardinality} in absolute accuracy.

\paragraph{CS-FLEURS subset analysis}
Table~\ref{tab:csfleurs_subset} shows that performance strongly depends on the CS-FLEURS subset. On seen synthetic pairs, most models trained with CS-FLEURS perform well. For example, H2, H3, S1-Top2, \add{M1-Top2, M2-Top2,} and G1 all reach around 99.6--99.8\% on XTTS1 seen pairs and MMS seen pairs. The main differences appear on unseen pairs and human-read speech.
On MMS unseen pairs, G1 achieves 28.5\%, whereas H2 obtains 0.0\%. This indicates that language-token set generation can represent unseen combinations of known \del{indivisual} \add{individual} languages in some synthetic conditions. \del{S1-Top2 achieves an even higher score of 56.7\% on the same subset, showing that soft-target language scoring is a very strong ranking-based baseline.} \add{However, M1-Top2 achieves 87.3\% on the same subset, showing that the sigmoid classifier is a very strong oracle Top-\(k\) baseline.}
In contrast, the near-zero accuracy across all models on XTTS2 unseen pairs demonstrates that generalizing from English-involved training data to non-English CS pairs remains an open challenge. \add{For the CS-trained models, almost all predictions on this subset include English, even though no XTTS2 pair contains it.}
The domain shift from synthetic training data to human-read speech makes the Read subset challenging. Specifically, H2 and G1 achieve near-zero accuracy, whereas \del{S1-Top2's 82.7\% accuracy on seen pairs suggests soft-target multi-class classifier is more robust to this synthetic-to-real domain gap.}\add{S1-Top2, M1-Top2, and M2-Top2 reach 82.7–88.6\% on seen pairs, suggesting that they are more robust to this synthetic-to-real domain gap. Read unseen pairs remain difficult, although M2-Top2 reaches 19.0\%.}

\paragraph{Effect of adding CS-YODAS as training data}
Adding CS-YODAS yields only limited improvement on CS-FLEURS: Read seen accuracy improves from 0.2\% to 0.8\% for the hard-target classifier (H2 vs.\ H3) and from 0.0\% to 0.2\% for the \del{multi-class}\add{multi-label} set generator (G1 vs.\ G2). However, these gains are small and do not solve the Read condition. Moreover, G2 loses the MMS unseen improvement observed in G1\add{, and the soft-target classifier degrades markedly (S1 vs.\ S2, 52.0\% to 33.0\% overall)}. \add{The one substantial gain is for the sigmoid classifier on Read unseen pairs, from 1.2\% to 19.0\% (M1 vs.\ M2).} \add{All comparisons in this paragraph are significant ($p<0.01$) under a two-sided exact McNemar test over paired utterances, with Holm correction across all 44 model-pair-by-subset tests we ran.}

This limited benefit is likely due to the mismatch between CS-YODAS and CS-FLEURS. CS-YODAS consists of long in-the-wild utterances of around 60 seconds, whereas CS-FLEURS utterances are much shorter, around 10 seconds. \add{CS-YODAS is also noisy, since it is mined from YouTube without studio conditions, whereas CS-FLEURS is read or synthesized.} This length\add{, noise,} and domain mismatch may make naive mixing of CS-YODAS with FLEURS and CS-FLEURS ineffective under the current training setup.

\begin{figure*}[htbp]
    \centering
    \includegraphics[width=0.85\linewidth]{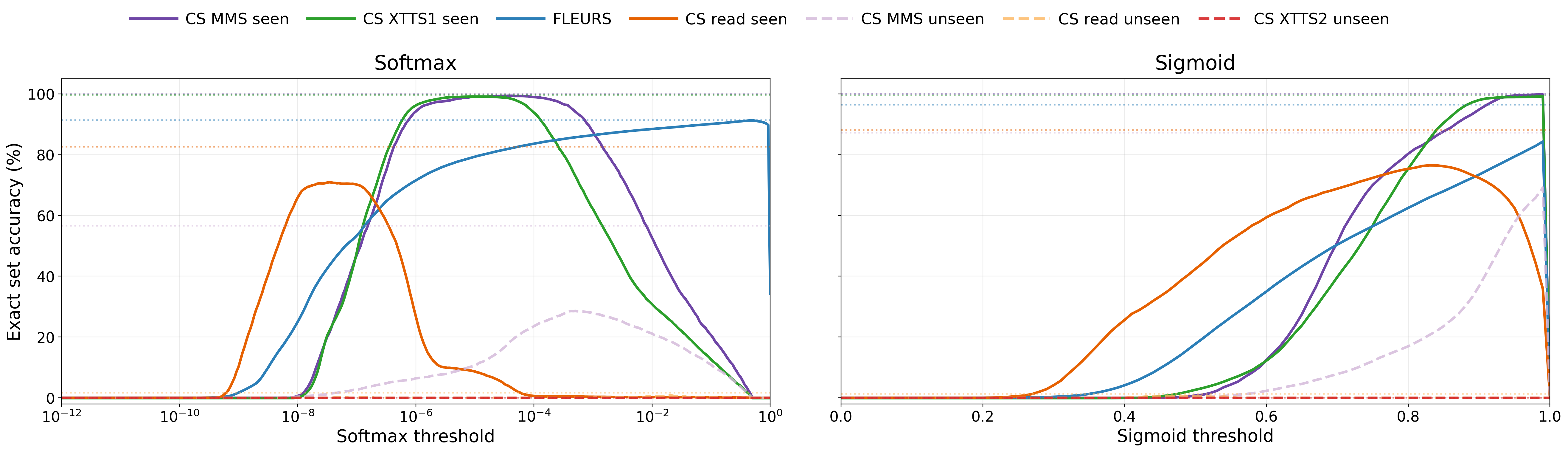}
    \caption{Threshold sensitivity of the soft-target (left) and sigmoid (right)
classifiers trained on FLEURS and CS-FLEURS (S1 and M1). Curves show exact set
accuracy obtained by thresholding softmax probabilities or sigmoid scores; dotted
lines show the corresponding oracle Top-\(k\) accuracies. No single threshold
recovers Top-\(k\) accuracy across subsets.}
\label{fig:threshold_sweep}
~\vspace{-2mm}
\end{figure*}

\paragraph{Prediction cardinality}
Table~\ref{tab:pred_cardinality} compares the four models that directly decide whether a CS-FLEURS utterance is predicted as one-language or two-language. For H2, this is determined by whether the predicted class is an individual language or an atomic pair. For G1, this is determined by the number of unique language tokens generated by the decoder.

Both H2 and G1 almost always predict one language on the Read subset, which explains their near-zero exact set accuracy there. In contrast, both models mostly predict two languages on synthetic seen subsets such as XTTS1 and MMS seen. On MMS unseen pairs, G1 predicts two languages for 96.8\% of utterances, whereas H2 does so for only 38.5\%. This partly explains why G1 improves over H2 on MMS unseen pairs. However, G1's exact set accuracy is still 28.5\%, showing that correct cardinality alone is not sufficient; the model must also identify the correct constituent languages.

\paragraph{Threshold sensitivity\del{ of the soft-target classifier}}
\label{sec:threshould}
Figure~\ref{fig:threshold_sweep} shows that exact set accuracy changes substantially as the threshold varies, and that the best threshold differs across subsets.

These results indicate that threshold-based decoding is not a stable replacement for oracle Top-\(k\) decoding. \del{Figure~\ref{fig:threshold_sweep} shows that detecting} \add{For the soft-target classifier, detecting} the second language of CS utterances requires a very small threshold (below $10^{-4}$, and around $10^{-8}$ for the Read subset), indicating that the model assigns only a tiny probability mass to the second language. In this regime, however, the second language of a CS utterance becomes indistinguishable from the spurious runner-up probabilities that the model also produces on monolingual speech: FLEURS accuracy keeps improving as the threshold increases toward $10^{-1}$. Since the operating ranges for CS and monolingual utterances barely overlap, no single global threshold can reliably handle both. \add{The sigmoid classifier keeps all subsets within a much narrower range, but its best FLEURS accuracy is about ten points below Top-1, and no single threshold is simultaneously best for FLEURS and the Read subset.}

\paragraph{Effect of English proportion}
Table~\ref{tab:miami_eng_ratio} shows the results on the Bangor Miami English--Spanish subset for the \del{three}\add{four} models  \del{(H2, S1, and G1)}trained on FLEURS and CS-FLEURS \add{(H2, S1, M1, and G1)}, grouped by English proportion. H2 and G1 fail almost completely across all groups, reaffirming that generalizing to real-world CS speech remains a challenge. \add{M1-Top2 is the strongest model in every group.} For \del{S1-Top2,}\add{both score-based classifiers, however,} accuracy degrades as the English proportion increases. This is likely because \add{English appears only as the embedded language in CS-FLEURS, so }the \del{CS-FLEURS }training data consists entirely of CS utterances with a low proportion of English.
These results highlight the difficulty of generalizing to CS utterances dominated by English.

\paragraph{Three-language code-switching}
We also evaluate the models\del{(H2, S1, and G1)} \add{trained on FLEURS and CS-FLEURS (H2, S1, M1, and G1)} on a subset of the Multilingual Soap Opera Corpus that features code-switching among three languages: English, Xhosa, and Zulu. \del{All three models, H2, S1-Top2, and G1, obtain 0.0\% unordered exact set accuracy.} \add{All four models obtain 0.0\% exact set accuracy.} \del{This indicates that none of the current models generalizes to three-language code-switching, and extending utterance-level CS-LID beyond two-language mixtures remains an open challenge.}\add{Extending utterance-level CS-LID beyond two-language mixtures thus remains an open challenge.}
\begin{table}[t]
\centering
\caption{Exact set accuracy (\%) on the Bangor Miami English--Spanish subset, grouped by English proportion. $\dagger$ marks oracle Top-2. Bold marks row maxima.}
\label{tab:miami_eng_ratio}
\vspace{-2mm}
\scriptsize
\setlength{\tabcolsep}{2.2pt}
\renewcommand{\arraystretch}{1.10}
\begin{tabular}{@{}lcccc@{}}
\hline
\textbf{English proportion} & \textbf{H2} & \textbf{S1-Top2}$^\dagger$ & \textbf{M1-Top2}$^\dagger$ & \textbf{G1} \\
\hline
All        & 0.18 & 16.46 & \textbf{26.16} & 0.07 \\
\hline
0--10\%    & 0.00 & 57.83 & \textbf{70.87} & 0.00 \\
10--20\%   & 0.25 & 28.26 & \textbf{43.49} & 0.25 \\
20--30\%   & 0.00 & 20.65 & \textbf{32.59} & 0.00 \\
30--40\%   & 0.39 & 15.69 & \textbf{24.71} & 0.00 \\
40--50\%   & 0.00 & 17.65 & \textbf{25.88} & 0.00 \\
50--60\%   & 0.31 & 11.49 & \textbf{17.08} & 0.00 \\
60--70\%   & 0.34 & 4.48  & \textbf{12.76} & 0.34 \\
70--80\%   & 0.00 & 4.40  & \textbf{14.80} & 0.00 \\
80--90\%   & 0.26 & 0.79  & \textbf{6.02}  & 0.00 \\
90--100\%  & 0.00 & 2.21  & \textbf{10.29} & 0.00 \\
\hline
\end{tabular}
\vspace{-4mm}
\end{table}

\section{Discussion}
Most CS-trained models are near-perfect on seen synthetic CS, but each fails once its assumption breaks: atomic-pair classes cannot represent unseen pairs, thresholding cannot separate CS from monolingual speech, and transfer from TTS-based training data to human-read speech remains challenging.
The cardinality analysis shows that predicting \emph{how many} and \emph{which} languages are separable subproblems. The set generator solves the former on MMS unseen pairs (96.8\%) but identifies the constituent languages
only 28.5\% of the time; the \del{soft-target classifier does}\add{score-based classifiers do} the reverse, externalizing cardinality to an oracle \(k\).
The synthetic-to-real gap remains the most pressing issue: only the \del{soft-target classifier}\add{score-based classifiers} with oracle Top-\(k\) \del{transfers}\add{transfer} to human-read seen pairs, while all models struggle on unseen pairs and real-world CS speech, and naively adding CS-YODAS does not consistently help. Better-matched CS training data may matter as much as further architectural changes.

\section{Conclusion}
We formulated utterance-level CS-LID as multi-label language-set prediction and compared our set generator against \del{two baselines, atomic-pair classification and soft-target classification}\add{atomic-pair classification and two score-based classifiers}, on a shared MMS-1B frontend. Oracle Top-\(k\) is the strongest baseline but requires the language count, and thresholding is not a viable substitute. Our set generator removes both assumptions but does not yet match Top-\(k\) accuracy. All approaches struggle on unseen-pair, English-dominant, and three-language CS speech, leaving robust set prediction for real-world CS an open challenge.

\section*{Acknowledgment}
This work was supported by JSPS KAKENHI Grant Numbers JP24H00742 and JP24H00748.

\newpage
\bibliographystyle{IEEEtran}
\bibliography{mybib}
\end{document}